\documentclass{article}
\newcommand{\al}{\alpha}
\newcommand{\bet}{\beta}
\newcommand{\pa}{\partial}

\newcommand{\si}{\sigma}

\newcommand{\tsi}{\tilde \si}

\newcommand{\ta}{\tau}

\newcommand{\om}{\omega}

\newcommand{\De}{\Delta}

\newcommand{\rar}{\rightarrow}

\newcommand{\non}{\nonumber}
\begin{document}
\title{Solvability of $F_4$ quantum integrable systems\footnote{Talk
    given by JCLV at the 12th INTERNATIONAL COLLOQUIUM Quantum Group
    and Integrable Systems, Prague, 12-14 June, 2003}}
\author{
 Juan C. Lopez Vieyra \footnote{vieyra@nuclecu.unam.mx} and
 Alexander V.~Turbiner \footnote{ turbiner@nuclecu.unam.mx}
\footnote{On leave of absence from the Institute for Theoretical
and Experimental Physics,
 Moscow 117259, Russia.} \\
{\small Instituto de Ciencias Nucleares, UNAM, A.P. 70-543, 04510 M\'exico
D.F., M\'exico}}
\date{}
\maketitle

\begin{abstract}
  It is shown that the $F_4$ rational and trigonometric integrable
  systems are exactly-solvable for {\it arbitrary} values of the
  coupling constants. Their spectra are found explicitly while
  eigenfunctions are obtained by pure algebraic means. For both
  systems new variables are introduced in which the Hamiltonian has an
  algebraic form being also (block)-triangular. These variables are
  a certain invariants of the $F_4$ Weyl group.  Both
  Hamiltonians preserve the same (minimal) flag of spaces of
  polynomials, which is found explicitly.\\[10pt]
{\small pacs: {02.30.Ik, 02.20.Sv, 02.70.Hm}}\\
{\small keywords: Quantum Integrable systems, Exact solvability, Root
  space invariants}

\end{abstract}

\section{Introduction}     

The $F_4$ rational and trigonometric models (see e.g. \cite{blt})
belong to a family of the Hamiltonian systems originally found in
the Hamiltonian reduction method introduced in the late seventies
by Olshanetsky and Perelomov \cite{Olshanetsky:1983}. Their
structure is related to the root systems of the $F_4$ algebra.
There exist both classical and quantum $F_4$ systems and they are
completely integrable.

In general, the Olshanetsky-Perelomov Hamiltonians possess
different symmetries (permutations, translation-invariance,
reflections, periodicity etc.). The idea of our approach is to
code all these symmetries in new coordinates and hence study
operators which can be called 'premature' and which give rise to
these Hamiltonians. It was formulated for the first time in
\cite{Ruhl:1995}, where $A_n$ rational and trigonometric models
(or, in other words, Calogero and Sutherland models) were studied.
Later this approach was developed for $G_2$ rational and
trigonometric models \cite{Rosenbaum:1998}. It is almost evident
that above-mentioned variables are nothing but a certain
Weyl-invariant functions with an extra property of periodicity in
trigonometric case. We will show that taking for $F_4$ quantum
Hamiltonians suitable invariants of the $F_4$ Weyl group as the
variables we arrive at the rational and trigonometric Hamiltonians
in an {\it algebraic} form, when the coefficient functions in
front of derivatives are polynomials. This property uncovers a
hidden algebra of both models and explains the exact solvability
of the $F_4$ quantum systems. Mostly, present Talk is based on the
article \cite{blt}. The results presented below appeared as a
conclusion of about five years of the intense study of the $F_4$
quantum systems.

\section{$F_4$ --Rational model}
The $F_4$-rational model describes a quantum particle in
four-dimensional space with a Hamiltonian given by
\begin{eqnarray}
\label{f4rath} H_{F_4}^{(r)} &=& \frac{1}{2} \sum_{i=1}^{4} \left(
-\pa_i^2 + \om^2 x_i^2 \right) + g\ \sum^4_{j>i} \left(
\frac{1}{(x_i-x_j)^2} +
\frac{1}{(x_i+x_j)^2} \right)\non \\
&&\hspace{-55pt} +\, g_1\  \sum_{i=1}^{4}\frac{1}{x_i^2} + 4g_1\
\sum^4_{ \nu's=0,1} \frac{1}{ \big[ x_1 + (-1)^{\nu_2}x_2+
(-1)^{\nu_3}x_3+ (-1)^{\nu_4}x_4 \big]^2} \ .
\end{eqnarray}
The form of the interactions is dictated by the set of positive
simple roots of the $F_4$ algebra.  This model contains two
independent coupling constants, $g$ and $g_1$, each of them
associated with roots of different lengths, and $\om$ is the
frequency of the harmonic oscillator term. The configuration space
coincides to the $F_4$ Weyl chamber.

Now we describe the procedure of solving the spectral problem
$H_{F_4}^{(r)} \psi=E\psi$ (see \cite{blt}).
\begin{enumerate}
\item At first let us remind the known form of the ground state
eigenfunction
\begin{eqnarray}
\Psi_{0} &=&
 \prod_{j<i} (x_i+ x_j)^{\nu} \prod_{j<i} (x_i- x_j)^{\nu}
\prod_{i} x_i^{\mu} \cdot \\ &&
\cdot  \prod^4_{\nu's=0,1} \hspace{-10pt} \big(x_1 +
(-1)^{-\nu_2}x_2 + (-1)^{-\nu_3}x_3+ (-1)^{-\nu_4}x_4 \big)^{\mu}
 e^{\displaystyle -\frac{\omega}{2} \sum x_i^2},  \nonumber
\end{eqnarray}
where $g = \nu(\nu - 1) > -\frac{1}{4},$ and $\ g_1=
\frac{1}{2}\mu (\mu -1) > -\frac{1}{8} \ .  $

 \item A general feature of the quantum rational Hamiltonians
stemming from the Hamiltonian reduction method is that all
eigenfunctions admit a factorization in a form of the ground state
eigenfunction $\Psi_0$ multiplied by a polynomial in Cartesian
coordinates. Exploiting it we gauge away the ground state,
\begin{equation}
 \label{f4rh}
h_{{F}_4}^{(r)} = 2(\Psi_{0})^{-1}\, {H}_{F_4}^{(r)}\, \Psi_{0} \ .
\end{equation}

 \item In order to try to find an algebraic form of the gauge
rotated Hamiltonian (\ref{f4rh}) we make a change of variables
\begin{equation}
 \label{chvr}
  (x_1,x_2,x_3,x_4) \rightarrow \big( {t}_1,\
 {t}_3,\ {t}_4,\ {t}_6\big)
\end{equation}
where the new variables $t_i,\ i=1,3,4,6$ are the Weyl invariant
polynomials with respect to the $F_4$ group of the lowest degrees
(in our notations, the indices $i=1,3,4,6$ correspond to the
degrees $2,6,8,12$).

The appropriate set of variables in which the problem takes an
algebraic form is
  \begin{eqnarray}
    \label{varrat}
 t_1 &=& \si_1, \qquad
 t_3 =  {\si}_3 - {\displaystyle \frac{1}{6}} {\si}_1\,{\si}_2\ ,\qquad
 t_4 = {\si}_4 -
 {\displaystyle \frac{1}{4}} \, {{\si}_1}\,{{\si}_3} +
 {\displaystyle \frac{1}{12}} \, {{\si}_2}^{2}, \non \\
 t_6 &=& {{\si}_4}\,{{\si}_2} - {\displaystyle \frac{1}{36}}\,{\si}_2^3
 - {\displaystyle \frac{3}{8}} {\si}_3^2 + {\displaystyle \frac{1}{8}}
 {\si}_1 {\si}_2 {\si}_3 - {\displaystyle \frac{3}{8}}{\si}_1^2 {\si}_4\ ,
 \end{eqnarray}
where $\si_n =\si_n(x^2)$ are the elementary symmetric polynomials
in variables $x_i^2,\ i=1,2,3,4$. In these variables the
Hamiltonian becomes a second order differential operator
\begin{equation}
\label{hf4ralg}
 {h_{F_4}^{(r)}}= \, \sum_{i \leq j}^6 {\cal A}_{i\,j}
 \frac{\pa^2}{\pa{t}_i\pa{t}_j} + \,\sum_{i=1}^6 {\cal B}_i
 \frac{\pa\ }{\pa{t}_i} \ ,
\end{equation}
with polynomial coefficients ${\cal A}_{ij} (={\cal A}_{ji}),\
{\cal B}_i$ in variables $t_{1,3,4,6}$ of degrees $\le 3$.
Therefore, the operator $h_{F_4}^{(r)}$ is in the algebraic form
we were looking for. Explicitly, the coefficients are
\[
{\cal A}_{1\,1} =  2\,t_1\ , \quad {{\cal A}_{1\,3}} = 6\,{{
{t}}_{3}}\ ,  \quad {{\cal A}_{1\,4}} = 8\,{{ {t}}_{4}}\ , \quad
{{\cal A}_{1\,6}} =  12\,{{ {t}}_{6}}\ ,
\]
\[
{{\cal A}_{3\,3}} =  - \frac{1}{3}\,{t}_3 \,{t}_{1}^2 +
\frac{10}{3}\,{t}_4\,{t}_1\ , \quad {{\cal A}_{3\,4}} =  -
{\displaystyle \frac {2}{3}} \,{{t}_{1}}^{2}\, {{t}_{4}} +
4\,{{t}_{6}}\ , \quad {{\cal A}_{3\,6}} = 8\,{{t}_{4}}^{2} -
\,{{t}_{1}}^{2}\,{{t}_{6}}\ ,
\]
\begin{equation}
\label{cof4r} {{\cal A}_{4\,4}} = - 2\,{{t}_{3}}\,{{t}_{4}} -
\,{{t}_{1}}\,{{t}_{6}}\ , \quad {{\cal A}_{4\,6}} = -
2\,{{t}_{1}}\,{{t}_{4}}^{2} - 3 \,{{t}_{3}}\,{{t}_{6}}\ ,
\end{equation}
\[
 {\cal B}_1 = 2\om{t}_1 + 24(\nu + \mu +\frac{1}{6})\ , \quad
 {\cal B}_3 = 6\om{t}_3 -2(\nu + \frac{\mu}{2}
+\frac{1}{4}){t}_1^{2}\ ,
\]
\[
 {\cal B}_4 =  8\om {t}_4 - 6(\nu + \frac{1}{3}){t}_3\ , \quad
 {\cal B}_6 =  12\om {t}_6 - 6(\nu + \frac{2}{3}){t}_1{t}_4\ .
\]
\end{enumerate}

\smallskip

It is easy to check that the Hamiltonian $ h_{\rm F_4}^{(r)}$ has
infinitely-many finite-dimensional {\it invariant} subspaces
\begin{equation}
\label{minflag}
 {\cal P}_{n}^{(F_4)} \ = \
\langle  {t}_1^{p_1} {t}_3^{p_3}
{t}_4^{p_4} {t}_6^{p_6}
| \ 0 \leq p_1 + 2 p_3 + 2 p_4 + 3
p_6 \leq n \rangle\ , n=0,1,\ldots
\end{equation}
with the characteristic vector $ \vec f \ =\ (1,2,2,3)$ formed
from the weight factors in front of $p_1,p_3,p_4,p_6$ in the
definition (\ref{minflag}).  These invariant subspaces form an
infinite {\it Flag}
\[
{\cal P}_0^{(F_4)} \subset  {\cal P}_1^{(F_4)} \dots \subset {\cal
P}_n^{(F_4)} \subset \dots
\]
The operator $h_{\rm F_4}^{(r)}$ with respect to action on
monomials ${t}_1^{p_1} {t}_3^{p_3} {t}_4^{p_4} {t}_6^{p_6}$ has
upper triangular form. The energies are given by
\[
  E_{p_1,p_2,p_3,p_4}= 2\, (p_1 + 3 p_2 + 4 p_3+ 6 p_4 + 2+12\mu+
  12\nu)\, \om\ ,  \quad p_i=0,1,2,\ldots
\]
in agreement with general formula given in
\cite{Olshanetsky:1983}. The spectrum does not depend on the
coupling constants $g$, $g_1$ (except for the reference point of
the energy), it is equidistant and it coincides with the
4-dimensional harmonic oscillator spectrum but with different
degeneracy, $n=p_1 + 3 p_2 + 4 p_3+ 6 p_4$. Since $h_{\rm
F_4}^{(r)}$ preserves the flag the calculation of eigenfunctions
is a linear-algebra procedure. It is important to make the
following remark. Let us define a general flag of spaces of
polynomials made out of
\[
 {\cal P}_{n} \ = \
\langle  {t}_1^{p_1} {t}_3^{p_3} {t}_4^{p_4} {t}_6^{p_6} | \ 0
\leq p_1 + \alpha_3 p_3 + \alpha_4 p_4 + \alpha_6 p_6 \leq n \rangle, \
n=0,1,\ldots
\]
with the characteristic vector $ \vec f \ =\
(1,\alpha_3,\alpha_4,\alpha_6)$. It can be easily checked that the
same operator written in different variables can preserve
different flags. Among these flags it can exist some minimal flag.
We call the flag {\it minimal} if the $\al$'s are minimal. Now we
proceed to search of the minimal flag if exists.

It is evident that the Weyl-invariant polynomials of fixed degrees
$t_{1,3,4,6}$ are defined ambiguously, up to invariants of lower
degrees
\begin{eqnarray*}
 t_1 &\rar&  t_1 \ , \non \\
 t_3 &\rar&  t_3 + A\, t_1^3 \ , \non \\
 t_4 &\rar&  t_4 + B_1\, t_1^4 + B_2\, t_1 t_3 \ ,\non \\
 t_6 &\rar&  t_6 + C_1\, t_1^6+  C_2\, t_1^3 t_3 +
 C_3\, t_1^2 t_4 + C_4\, t_3^2 \ ,
\end{eqnarray*}
where $A, B, C$ are arbitrary numbers. Exploiting this ambiguity
we were able to find {\it many} different algebraic forms for the
$F_4$-rational Hamiltonian $h_{\rm F_4}^{(r)}$ which preserve
different flags of polynomials. Here we present a partial list of
characteristic vectors of such flags for the $F_4$-rational model
\[
\begin{array}{ccccc}
(1,2,2,3) & (1,3,3,5) & (1,5,5,8)  &(1,6,6,9)  &(1,6,7,10) \\
(1,2,3,4) & (1,4,4,6) & (1,5,5,9)  &(1,6,6,10) &(1,7,7,11) \\
(1,2,3,5) & (1,4,4,7) & (1,5,7,9)  &(1,6,6,11) & \ldots \\
\end{array}
\]
By comparison of the characteristic vectors it is easy to see that
the minimal flag exists and is characterized by the vector
$(1,2,2,3)$. This result is in variance with one obtained in
\cite{Ruehl:1998} where it was stated that the minimal vector is
$(1,2,3,5)$. We were able to 'trigonometrize' the variables
leading to the minimal flag, getting them as a rational limit of a
certain trigonometric Weyl-invariant functions for those the $F_4$
trigonometric Hamiltonian appears in an algebraic form (see
below). It seems that this feature holds for other models.

\section{$F_4$ --Trigonometric model}
The Hamiltonian of the $F_4$ trigonometric model describes a
four-dimensional quantum system in a periodic potential in all
four directions. It is given by
 \begin{eqnarray}
 \label{hf4t}
  {\cal H}_{\rm F_4}^{(t)}(x) \ &=&\ -\frac{1}{2} \sum_{i=1}^{4}
 \pa_{x_i}^2 + 2g \beta^2 \sum_{j>i}^4 \left(
\frac{1}{\sin^2 \beta(x_i-x_j)} + \frac{1}{\sin^2 \beta(x_i+x_j)}
\right) \\ && \hspace{-30pt} +
 2g_1 \beta^2 \left(
 \sum_{i=1}^{4}\frac{1}{\sin^2 2\beta{x_i}} +
 \sum_{ \nu's=0,1}^4 \frac{1}{\sin^2 \beta \left[ x_1 +
(-1)^{\nu_2}x_2+ (-1)^{\nu_3}x_3+ (-1)^{\nu_4}x_4 \right]}
\right)\ . \non
 \end{eqnarray}
where the parameter $1/\beta$ has meaning of the period, $g,g_1$
are the coupling constants associated with roots of different
length. Configuration space of the problem coincides with the
$F_4$ Weyl alcove.

The corresponding ground state eigenfunction is given by
\[
 \Psi_0^{(t)} (x,\beta) =
 \left\{ \De_+(x,\beta) \De_-(x,\beta) \right\}^{\nu}
 \left\{\Delta_0 (x,2\beta) \Delta (x,2\beta) \right\}^{\mu} \ ,
\]
where $g=\nu (\nu-1)/2\ ,\ g_1=\mu (\mu-1)$, and
\[
 \De_{\pm}(x,\beta) \ =
 \beta^{-6}\prod^4_{j<i}\sin\beta(x_i\pm x_j), \quad
 \De_{0}(x,2\beta) \ =\
 \beta^{-4}\prod^4_{i=1}\sin 2\beta x_i\ ,
\]
\[
 \De (x,2\beta) \ =
 \beta^{-8}\prod^4_{\nu's=0,1}\sin \beta\left[x_1 +
 (-1)^{\nu_2}x_2 + (-1)^{\nu_3}x_3+ (-1)^{\nu_4}x_4 \right]\ ,
\]
are trigonometric analogues of the Weyl determinants.

In order to solve the spectral problem we again gauge away the
ground state and define the gauge-rotated Hamiltonian
\[
 h_{\rm F_4}^{(t)} \ =\ -2\big(\Psi_{0}^{(t)}(x)\big)^{-1}({\cal
 H}_{\rm F_4}^{(t)}-E_0) \big(\Psi_{0}^{(t)}(x)\big)\ ,
\]
Now we search for a change of variables
\[
 (x_1,x_2,x_3,x_4) \rar \big(\tau_1,\tau_3,\tau_4,\tau_6 \big)\ ,
\]
in which the gauge-rotated Hamiltonian $h_{\rm F_4}^{(t)}$ may
take an algebraic form. Similarly to what was done for the
rational case we look for the Weyl-invariant functions with an
extra property of periodicity in each $x$-direction. The solution
we found \cite{blt} is surprisingly easy and is given by
\[
\begin{array}{l}
 \tau_{1} = \,\tsi_{1}-\frac{2\beta^2}{3}\tsi_{2}\ , \\[3pt]
 \tau_{3} = \,\tsi_3 - \frac{1}{6}\,\tsi_1\,\tsi_2-
2\beta^2(\tsi_{4}-\frac{1}{36}\tsi_{2}^2)\ , \\[3pt]
 \tau_{4} = \,\tsi_4
- {\frac{1}{4}}\,\tsi_1\,\tsi_3+ {\frac{1}{12}}\,\tsi_2^{2}, \\[3pt]
 \tau_{6} = \,\tsi_4\,\tsi_2 - {\frac{1 }{36}}\,\tsi_2^{3}-
 {\frac{3}{8}}\,\tsi_3^{2}+\frac{1}{8}
\,\tsi_1\,\tsi_2\,\tsi_3 - \frac{3}{8}\,\tsi_1^{2}\,\tsi_4\ ,\\[4pt]
\end{array}
\]
(cf. (\ref{varrat})) where $\tsi_n =\si_n(\frac{\sin^2(\beta x
)}{\beta^2})$ are elementary symmetric polynomials of the periodic
arguments. It is clear that $\tau_{1,3,4,6}$ are the Weyl
invariant trigonometric (periodic) variables. In these variables
the Hamiltonian becomes a second order differential operator with
polynomial coefficients
\[
 h_{\rm F_4}^{(t)}
= \, \sum_{a<b} A_{ab} \frac{\pa^2}{\pa\tau_a\pa\tau_b}\ +\,
  \sum_{a} \left(B_a + C_a\right)\frac{\pa\ } {\pa\tau_a} \ ,
 \qquad a,b = 1,3,4,6. \ ,
\]
where
\begin{eqnarray*}
 A_{11} &=& 4\,\tau_{1}-4\beta^2{
\tau}_{1}^2-\frac{32}{3}\beta^4\tau_{3}-\frac{128}{9}\beta^6
\tau_{4}, \qquad \quad
\\
 A_{13} &=& 12\,\tau_{3}-\frac{8}{3}\beta^2(4\tau_{1}\tau_{3}+\tau_{4})-
\frac{32}{9}\beta^4\tau_{1}\tau_{4},
\\
 A_{14} &=&  16\,\tau_{4}
-\frac{40}{3}\beta^2\tau_{1}\tau_{4}- \frac{16}{3}\beta^4\tau_{6},
\\
 A_{16} &=&   24\,\tau_{6}
- 20\beta^2\tau_{1}\tau_{6}- \frac{32}{3}\beta^4\tau_{4}^2,
\\
A_{33} &=&  -\frac {2}{3} \,\tau_{1}^{2}\,\tau_{3} +\frac {20}{3}
\,\tau_{1}\,\tau_{4} - \frac{8}{9}\beta^2\,(18\tau_{3}^2
+\tau_{1}^2\,\tau_{4}
 +12\tau_{6}),
\\
A_{34} &=& -\frac{4}{3} \,\tau_{1}^{2}\, \tau_{4} + 8\,\tau_{6}
 -\frac{4}{3}\beta^2\,(\tau_{1}\,\tau_{6}
 +12\tau_{3}\,\tau_{4}),
\\
A_{36} &=& 16\,\tau_{4}^{2} - 2\,\tau_{1}^{2}\,\tau_{6}
 -\frac{8}{3}\beta^2 (9\tau_{3}\,\tau_{6}
 +\tau_{1}\,\tau_{4}^2),
\\
A_{44} &=& - 4\,\tau_{3}\,\tau_{4} -
2\,\tau_{1}\,\tau_{6}-24\beta^2\tau_{4}^2,
\\
A_{46} &=& - 4\,\tau_{1}\,\tau_{4}^{2} - 6 \,\tau_{3}\,\tau_{6}-
36\beta^2 \tau_{4}\tau_{6},
\\
A_{66} &=& - 12\tau_3\tau_4^2 - 6 \tau_{1}\tau_{4}\tau_6 -
8\beta^2(6\tau_{6}^2 +\tau_{4}^3),
\\
 B_1 &=& 8-8\beta^2 \tau_1 , \hspace{45pt}
B_3 = - \tau_{1}^{2}- \frac{56}{3}\beta^2\tau_{3}
-\frac{32}{9}\beta^4\tau_{4},
\\
 B_4 &=& - 4\,\tau_{3}-\frac{88}{3}\beta^2\tau_4,
\quad \quad
 B_6 = -8 \tau_{1}\tau_{4}- 56\beta^2 \tau_6,
\\
 C_{1} &=& 48(\nu+\mu) -8\beta^2(5\nu+6\mu)
\tau_1
\\
 C_{3}&=&-2(2\nu+\mu)\tau_1^2-16\beta^2(3\nu+5\mu)\tau_3
 \\
 C_{4}&=& -12\nu\tau_3-24\beta^2 (3\nu+4\mu)\tau_4
\\
 C_{6} &=& -12\nu\tau_1\tau_4 -48\beta^2(2\nu+3\mu)\tau_6
\end{eqnarray*}
Similar to the rational case the coefficients $A_{ij}$ and $B_i,
C_i$ are polynomials in $\ta$'s of the degrees not higher than
three and two, correspondingly.

The operator $h_{F_4}^{(t)}$ presents the algebraic form of the
$H_{F_4}^{(t)}$ Hamiltonian. It preserves the {\bf same} flag of
spaces of polynomials ${\cal P}^{(F_4)}$ as in the rational case
$(1,2,2,3)$ (now in variables ${\tau}_{1,3,4,6}$). But it is {\bf
NOT} triangular with respect to the action on monomials
$\tau_1^{p_1}\tau_3^{p_3}\tau_4^{p_4}\tau_6^{p_6}$ in variance to
the general statement made in \cite{Khastgir:00}. Making a {\it
singular} in $\beta$ transformation preserving the flag

\[
\rho_1 = \tau_1 \ , \qquad
\rho_3 = \tau_3 - \frac{1}{8}\bet^{-2}\tau_1^2 \ ,
\]
\[
\rho_4 = \tau_4 - \frac {3}{16}\bet^{-4}\tau_1^2 \ , \quad
\rho_6 = \tau_6 - \frac{3}{4}\bet^{-2}\tau_1 \tau_4
 + \frac{3}{64}\bet^{-6}\tau_1^3 \ ,
\]
we arrive at the $F_4$ trigonometric Hamiltonian
$h_{F_4}^{(t)}(\rho)$ in triangular form. From this form we can
immediately calculate the spectrum:

\begin{eqnarray}
 E_{n} &=& 4 \, [p_1 (p_1 + 2 p_3 + 3 p_4 + 4 p_6) +
 2p_3(p_3+2p_4+3p_6)+p_4(3p_4+8p_6) +6p_6^2 + \non\\&& \nu (5 p_1 +
 6 p_3 + 9 p_4 + 12 p_6) + 2\mu (3 p_1 + 5 p_3 + 6 p_4 + 9 p_6)]\,
 \bet^2 + \non\\&& 4\bet^2 (7 \nu^2 + 14 \mu^2 + 18 \nu \mu)\ ,
 \qquad p_i=0,1,2\ldots \ , \non
\end{eqnarray}
 Since we know the flag of invariant subspaces, we can calculate
eigenfunctions by linear algebra means. It is worth mentioning
that in \cite{Ruehl:1999} it was found non-minimal flag
$(1,2,3,4)$.

\bigskip {\small {\bf Acknowledgments} \\ This work is supported in
part by the grant {\it IN124202} (UNAM).}

\end{document}